\input harvmac.tex

\def\cp{${\bf P}^1$}

\noblackbox
\lref\conmet{P. Candelas and X. de la Ossa, ``Comments on Conifolds,''
{\it Nucl. Phys.} {\bf B342} (1990) 246.}
\lref\berjoh{M. Bershadsky and A. Johansen, ``Colliding Singularities
in F-theory and Phase Transitions,'' hep-th/9610111.}
\lref\asty{O. Aharony, J. Sonnenschein, S. Theisen, and S. Yankielowicz, 
work in progress.}
\lref\sw{N. Seiberg and E. Witten, ``Electric-Magnetic Duality,
Monopole Condensation, and Confinement in ${\cal N}=2$ supersymmetric
Yang-Mills Theory,'' {\it Nucl. Phys.} {\bf B426} (1994) 19, 
hep-th/9407087.}
\lref\ksw{D. Kutasov, A. Schwimmer, and N. Seiberg, ``Chiral Rings,
Singularity Theory, and Electric-Magnetic Duality,'' 
{\it Nucl. Phys.} {\bf B459} (1996) 455, hep-th/9510222.}
\lref\intsei{K. Intriligator and N. Seiberg, ``Phases of 
${\cal N}=1$ Supersymmetric Gauge Theories in Four Dimensions,''
{\it Nucl. Phys.} {\bf B431} (1994) 551, hep-th/9408155.}
\lref\katz{S. Katz and C. Vafa, ``Matter from Geometry,'' hep-th/9606086.}
\lref\finite{
See for example: R. Leigh and M. Strassler, ``Exactly Marginal Operators
and Duality in Four-Dimensional ${\cal N}=1$ Supersymmetric Gauge Theory,''
{\it Nucl. Phys.} {\bf B447} (1995) 95, hep-th/9503121, and references
therein.}
\lref\review{For a review with references see: K. Intriligator and
N. Seiberg, ``Lectures on Supersymmetric Gauge Theories and
Electric-Magnetic Duality,'' hep-th/9509066.}
\lref\banks{T. Banks and A. Zaks, ``On the Phase Structure of 
Vectorlike Gauge Theories with Massless Fermions,``
Nucl. Phys. {\bf B196} (1982) 189.}
\lref\argdoug{P. Argyres and M. Douglas, ``New Phenomena in $SU(3)$ 
Supersymmetric Gauge Theory,'' {\it Nucl. Phys.} {\bf B448} (1995)
93, hep-th/9505062.}
\lref\swone{N. Seiberg and E. Witten, ``Electric-Magnetic Duality,
Monopole Condensation, and Confinement in ${\cal N}=2$ Supersymmetric 
Yang-Mills Theory,'' {\it Nucl. Phys.} {\bf B426} (1994) 19, 
hep-th/9407087.}
\lref\aspg{P. Aspinwall and M. Gross, ``The $SO(32)$ Heterotic String
on a $K3$ Surface,'' hep-th/9605131.}
\lref\swtwo{N. Seiberg and E. Witten, ``Monopoles, Duality, and
Chiral Symmetry Breaking in ${\cal N}=2$ Supersymmetric QCD,'' 
{\it Nucl. Phys.}
{\bf B431} (1994) 484, hep-th/9408099.}
\lref\ntwo{P. Argyres, R. Plesser, N. Seiberg, and E. Witten, 
``New ${\cal N}=2$ Superconformal Field Theories in Four Dimensions,''
hep-th/9511154.}
\lref\generaliz{ 
T. Eguchi, K. Hori, K. Ito, and S-K Yang, ``Study of ${\cal N}=2$ 
Superconformal
Field Theories in 4 Dimensions,'' hep-th/9603002\semi
T. Eguchi and K. Hori, ``${\cal N}=2$ Superconformal Field Theories in
4 Dimensions
and A-D-E Classification,'' hep-th/9607125.} 
\lref\mrdprobe{M. Douglas, ``Branes within Branes,'' hep-th/9512077\semi
M. Douglas and G. Moore, ``D-branes, Quivers, and ALE Instantons,''
hep-th/9603167\semi
M. Douglas, ``Gauge Fields and D-branes,'' hep-th/9604198\semi
M. Douglas, D. Kabat, P. Pouliot, and S. Shenker, ``D-branes 
and Short Distances in String Theory,'' hep-th/9608024. 
}
\lref\brprobe{
T. Banks, M. Douglas, and N. Seiberg, ``Probing F-theory with Branes,''
hep-th/9605199.} 
\lref\seiberg{
N. Seiberg, ``IR Dynamics on Branes and Space-Time Geometry,''
hep-th/9606017\semi   
K. Dasgupta and S. Mukhi, 
``F-theory at Constant Coupling,'' hep-th/9606044\semi
K. Intriligator and N. Seiberg, ``Mirror Symmetry in Three Dimensional
Gauge Theory,'' hep-th/9607027\semi 
O. Ganor, ``Toroidal Compactification of Heterotic 6-D Noncritical
Strings Down to Four-Dimensions,'' hep-th/9608109\semi 
N. Seiberg, ``Five Dimensional SUSY Field Theories, Non-trivial Fixed Points,
and String Dynamics,'' hep-th/9608111\semi
D.R. Morrison and N. Seiberg, ``Extremal Transitions and Five-Dimensional
Supersymmetric Field Theories,'' hep-th/9609069\semi
M.R. Douglas, S. Katz, and C. Vafa, ``Small Instantons, Del-Pezzo Surfaces,
and Type I$^\prime$ Theory,'' hep-th/9609071\semi
E. Witten, ``Physical Interpretation of Certain Strong Coupling
Singularities,'' hep-th/9609159\semi 
N. Seiberg, ``Non-trivial Fixed Points of the Renormalization Group in
Six Dimensions,'' hep-th/9609161. 
}
\lref\kss{S. Kachru, N. Seiberg, and E. Silverstein, ``SUSY Gauge Dynamics
and Singularities of 4d ${\cal N}=1$ String Vacua,'' hep-th/9605036.}
\lref\ftheory{C. Vafa, ``Evidence for F-theory,'' hep-th/9602022\semi 
D. Morrison and C. Vafa, ``Compactifications of F-theory on
Calabi-Yau Threefolds I,II,'' hep-th/9602114, hep-th/9603161.}
\lref\bikmsv{M. Bershadsky, K. Intriligator, S. Kachru, D. Morrison,
V. Sadov, and C. Vafa, ``Geometric Singularities and Enhanced Gauge
Symmetries,'' hep-th/9605200.} 
\lref\senetc{A. Sen, ``F-theory and Orientifolds,'' hep-th/9605150.} 
\lref\Esix{J. Minahan and D. Nemeschansky, ``An ${\cal N}=2$
Superconformal Fixed Point
with $E_6$ Global Symmetry,'' hep-th/9608047\semi
W. Lerche and N. Warner, ``Exceptional SW Geometry from ALE 
Fibrations,'' hep-th/9608183\semi
J. Minahan and D. Nemeschansky, ``Superconformal Fixed Points with $E_n$ 
Global Symmetry,'' hep-th/9610076.}

\Title{RU-96-96, hep-th/9610205}
{\vbox{\centerline{New ${\cal N}=1$ Superconformal Field Theories}
	\vskip4pt\centerline{in Four Dimensions from D-brane Probes}}}
\centerline{Ofer Aharony, Shamit Kachru, and Eva Silverstein 
\footnote{$^\star$}
{oferah, kachru, evas@physics.rutgers.edu 
}}
\bigskip\centerline{Department of Physics and Astronomy}
\centerline{Rutgers University}
\centerline{Piscataway, NJ 08855-0849}

\vskip .3in
We present several new examples of nontrivial 
4d ${\cal N}=1$ superconformal field theories.  Some of these theories
exhibit exotic global symmetries, including non-simply laced groups
(such as $F_4$). 
They are obtained by studying threebrane probes in 
F-theory compactifications on elliptic Calabi-Yau threefolds.
The geometry of the compactification encodes in a simple
way the behavior of the gauge coupling and the K\"ahler potential
on the Coulomb branch of these theories.  

\Date{10/96} 

\newsec{Introduction}

Recent developments have led to dramatic improvements in our 
understanding of
quantum field theory and string theory.
For example, inspired by the beautiful 
work of \refs{\mrdprobe,\senetc,\brprobe}, 
the study of quantum field theories on
brane probes has led to the discovery of a new and surprising 
set of exotic fixed points of the renormalization group
in three, four, five, and six dimensions \refs{\seiberg,\Esix}. 
For some of these fixed points no Lagrangian is known that flows to
them in the infrared, suggesting that other methods are necessary to
study general field theories (and, in particular, conformal field
theories). 
The theories studied in \refs{\seiberg,\Esix} have the equivalent of
4d ${\cal N}=2$ supersymmetry.
In this paper we present
a new class of renormalization group fixed points
with 4d ${\cal N}=1$ superconformal symmetry.
These are obtained as the effective worldvolume theory on
threebrane probes in six-dimensional ${\cal N}=1$ string compactifications.

The known examples of ${\cal N}=1$ superconformal theories fall into two 
classes.  One class consists of finite theories, in which the coupling
parametrizes a line of fixed points \finite. 
For these there is a Lagrangian description of the fixed point theory
(which is useful at least at weak coupling). One of the new
superconformal field theories we discuss is of this type.  
Another class which has been investigated in detail in the past two  
years consists of the infrared limit 
of theories like ${\cal N}=1$ $SU(N_c)$ gauge 
theories with ${3\over 2}N_c < N_f < 3 N_c$
flavors of quarks \review.  
There are two known asymptotically free quantum field
theories which flow to these nontrivial fixed points in the IR.   
These theories exist for a fixed (unknown) value of the IR gauge coupling.
Most of the 
theories we find exemplify a third class of ${\cal N}=1$ superconformal
theories: we find nontrivial fixed points 
with exotic global symmetries (including $F_4$) 
at fixed values of the coupling.
There are no known candidates for Lagrangian field theories which flow to
these theories in the infrared.  Under relevant perturbations, they flow
to $SU(2)$ or $U(1)$ gauge theories with known matter content.  This is
similar in many ways to the theories recently discussed in \seiberg.

The six-dimensional theories of interest are obtained
by compactification of the heterotic string on $K3$ with
$12+n$ instantons in one $E_8$ factor and $12-n$ in the
other.  This is dual \ftheory\ to F-theory compactified on the
Calabi-Yau manifold which is an
elliptic fibration over the Hirzebruch surface $F_n$.  
When the $K3$ on the heterotic side is an elliptic fibration over \cp,
we can consider a threebrane obtained by wrapping the
fivebrane on the $T^2$ fiber. 
The quantum field theory on this threebrane has an $SU(2)$
gauge symmetry with an adjoint scalar $X$ and an additional
scalar field which parametrizes the position on the base
$\bf{P^1}$. 
On
the F-theory side, this is dual to a Dirichlet threebrane probe
whose moduli space includes the $F_n$ surface.  The
threebrane is the natural probe to consider in F-theory,
since it is invariant under the $SL(2,{\bf Z})$ duality symmetry
of type IIB string theory.   
Viewing $F_n$ as a $\bf {P^{1}}$ bundle over $\bf{P^{1}}$, 
$\tr(X^2)$ of the 
heterotic theory maps to the position on the 
$\bf {P^1}$ fiber while the additional scalar maps to the
position on the base $\bf {P^1}$.
 
There is a simple relation between the spacetime theory and the 
worldvolume
quantum field theory of a D-brane probe. 
Fields in the spacetime theory become
parameters in the brane theory.  The spacetime gauge symmetry
which is apparent locally to the probe is (part of) the
global symmetry on the brane.  The complex structure
$\tau$ of the elliptic fiber in 
F-theory is the infrared gauge coupling on the threebrane probe.
The metric of the compactification manifold gives the
metric on the Coulomb branch of the field theory on the probe.  
This is part of a general story in which short distance effects
in spacetime are translated (via a relation between distances
and masses) to long distance effects on D-brane probes.
These facts allow us to identify nontrivial interacting ${\cal N}=1$
fixed points, which in some cases possess exceptional global
symmetry, by moving the probe to appropriate points in spacetime. 
For generic points on the base ${\bf P^1}$, the
models reduce in the infrared to 4d ${\cal N}=2$ theories
studied on threebrane probes in the eight-dimensional spacetime
theory \refs{\senetc,\brprobe}.  However, at special points on this
${\bf P^1}$ having to do with the nontrivial fibration,
the brane theory changes and exhibits properties 
indicating that it has become a nontrivial ${\cal N}=1$ superconformal theory. 
The consistency of our results suggests that it is valid
to use threebranes as probes in six dimensions, despite the
fact that they induce a deficit angle in the noncompact space. This
could be related to the fact that our analysis uses only local
properties of the Calabi-Yau manifold, and could in
principle be performed also when this manifold is not compact.

Compactifications of the heterotic string on $K3$ yield an 
intricate web of 6d ${\cal N}=1$ theories, the F-theory
description of which was worked out in \refs{\ftheory,\aspg,\bikmsv}.
Because
of the relations reviewed in the previous paragraph,
Higgsing in spacetime decreases the global symmetry on the brane.
This partially determines the operator content and couplings
present in the worldvolume theory.   
In section two, we review the relevant spacetime theories.
In section three, we discuss theories with
$SO(n)$ global symmetry, for which we have a Lagrangian description.  
Theories with exceptional global
symmetry, and other theories for which we do not have a Lagrangian
description, are covered in section four.  Finally, we present
our conclusions and possible generalizations in section five.

\newsec{Three-Brane Probes and Nontrivial Fixed Points}

A fairly detailed map from the moduli space of the $E_8 \times E_8$
heterotic string
on $K3$ with $(12+n, 12-n)$ instantons in the two $E_8$s to the 
complex moduli of the elliptic fibration over $F_n$ was given in 
\bikmsv.  Following the notation of that paper, we denote the coordinate
on the $\bf{P^1}$ base of the $F_n$ as $z_{2}$ and the coordinate on the
$\bf{P^1}$ fiber as $z_1$.   The unbroken subgroups $H_{1,2}$
of the two $E_8$
gauge groups of the heterotic theory are localized on the divisors at
$z_1=0$ and $z_1 = \infty$.  If we move the threebrane probe to either of these
divisors, we obtain (at least) an $H_{1,2}$ $\it global$ symmetry in
the long-distance quantum field theory on the probe.  

In the F-theory description, one obtains the unbroken spacetime
gauge symmetry by
fibering an $A-D-E$ singularity over the $z_2$ sphere.  
At generic points on the $z_2$ plane, the brane probe flows in
the infrared to an ${\cal N}=2$ superconformal theory with $A-D-E$ global
symmetry, and the additional spacetime sphere has no effect.
At special points
on the $z_2$ plane, the singularity becomes worse.  There are two
things which can happen at such special points: 

\noindent  (1) The $A-D-E$ gauge group undergoes an outer automorphism
as we go around the special point \aspg.  In this case, the gauge group is
broken from the simply laced $A-D-E$ group to a non-simply laced
group  
$Sp(k)$, $SO(2k+1)$, $F_4$, or $G_2$.  
At this type of special point, we will find that 
in some cases the probe theory flows instead
to an ${\cal N}=1$ superconformal theory with a 
non-simply laced global symmetry group.

\noindent (2) Charged matter of the spacetime
gauge group is localized at the special
point.  In this case the theory on the brane probe sees extra light 
particles as one approaches the special point.  If the light particles in
the probe theory have mutually non-local charges, one again obtains a 
nontrivial superconformal field theory.   We find evidence that in
cases where whole spacetime hypermultiplets are localized at a point
the brane probe sees an ${\cal N}=2$ superconformal 
theory in the infrared, while
at ``half hypermultiplet'' points it sees an ${\cal N}=1$ 
superconformal theory.   
Examples of both types of localization are 
known \refs{\ftheory,\aspg,\bikmsv}.
Note that in the cases where the spacetime gauge group is non-simply laced,
the matter is not necessarily
localized in this manner but there are interesting
points of type (1).   

Under a relevant perturbation, the new fixed points we find flow to
$SU(2)$ or $U(1)$ gauge theories with Coulomb phases (which correspond
to generic points on $D_N$ and $A_N$ singular curves).  The F-theory
description of the spacetime theory 
involves an elliptically fibered Calabi-Yau manifold defined by an
equation of the form
\eqn\weier{y^{2} = x^{3} + x f(z_{1},z_{2}) + g(z_{1},z_{2}). }
Expanding \weier\ around singular points 
directly provides the ``Seiberg-Witten curve'' determining
the gauge coupling function on the Coulomb branch for the field theory
on our brane probe \refs{\sw,\swtwo,\intsei}.  Note that our theories
always have Coulomb branches, since at a generic point on the $F_n$
the worldvolume theory on the brane probe is a $U(1)$ gauge theory.
The singular points in the elliptic fibration (near which the field theory on
the brane probe has interesting behavior) are points where 
the discriminant
\eqn\discdef{\Delta ~=~ 4f^{3} + 27 g^{2}}
vanishes. 
Then, using 
\eqn\j{j(\tau) \sim {f^3\over{\Delta}}}
which gives the complex structure of the fiber torus, 
we can determine the gauge coupling $\tau$ 
on the Coulomb branch around our new fixed points.

In the F-theory compactifications we are studying, one obtains chains
of different models related by Higgsing in spacetime.  This is reflected
in Diagram 1, which is borrowed from \bikmsv.  
\bigskip
$$\matrix{SU(3)_2 & \leftarrow &Sp(3) & \leftarrow & \, & SU(6)
&\leftarrow & SO(12) & \,& \,\cr
\, &\, & \, & \, & \, & \,  &\, & \downarrow  &
\,& \,\cr
\, &\, &\, & \, & \, & \downarrow  &\, & SO(11) &
\leftarrow & E_7\cr
\, &\, &\, & \, & \, & \,  &\, & \downarrow  &
\,& \downarrow\cr
\downarrow &\, &\downarrow & \, & \, & SU(5)  &\leftarrow &
SO(10) & \leftarrow & E_6\cr
\, &\, &\, & \, & \, & \,  &\, & \downarrow  & \,&
\downarrow\cr
\, &\, &\, & \, & \, & \,  &\, & SO(9)  &
\leftarrow & F_4\cr
\downarrow &\, &\downarrow & \, & \, & \downarrow  &\, &
\downarrow \cr
\, &\, &\, & \, & \, & \,  &\, & SO(8) \cr
\, &\, &\, & \, & \, & \,  &\, & \downarrow \cr
\downarrow &\, &Sp(2) & \leftarrow & \, & SU(4)  &
\leftarrow & SO(7) \cr
\, &\, &\, & \, & \, & \downarrow  &\, & \downarrow \cr
\, &\, &\downarrow & \, & \, & SU(3)  &\leftarrow & G_2 \cr
\, &\, &\, & \, & \, & \downarrow & \,\cr
SU(2)_2 &\leftarrow &SO(4)       & \rightarrow &
\, &SU(2) & \,\cr}$$
\bigskip
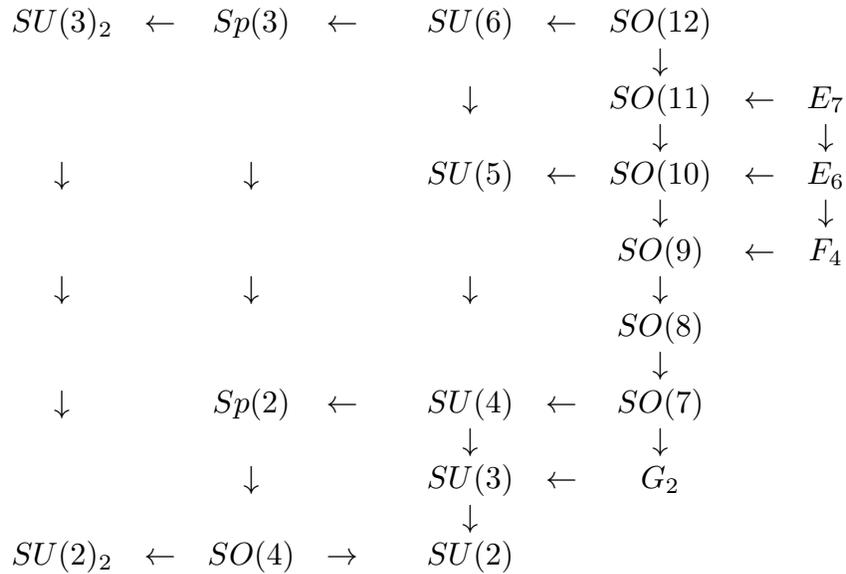
\centerline{Diagram 1: Phase Diagram}
\bigskip

\noindent
In the context of our
brane probe field theories, the $\it gauge$ symmetries in spacetime
become $\it global$ symmetries of the infrared theory on the probe.
The arrows (reflecting spacetime Higgsing) become flows between 
the worldvolume field theories under relevant perturbations.
When the Higgsing in spacetime breaking gauge group $G$ to gauge
group $H$ is accomplished by giving VEVs to fields in the
$R$ representation of $G$, we expect that 
these spacetime fields act as parameters in the worldvolume field theory
coupled to operators in the $\bar R$ representation.  Turning on these
parameters introduces a relevant perturbation under which the theory
flows from the fixed point with global symmetry containing $G$ 
to a new one with global symmetry containing $H$ in the infrared.

\newsec{Lagrangian Field Theories and Fixed Lines 
from $D_N$ Singularities}

\subsec{$SO(2N)$ Theories}

In order to illustrate our methods, it is useful to begin with
examples for which we can propose a Lagrangian description.  Most of
these theories flow in the IR to free theories, but they may reached by
relevant perturbations from our nontrivial fixed points.

Let us begin with a $D_N$ singularity in the ``upper''
\cp\ ($N \ge 4$). 
According to \brprobe, the field theory of a threebrane probe near
such a singularity is an ${\cal N}=2$ supersymmetric
$SU(2)$ gauge theory with $N$ quark flavors, which 
has an $SO(2N)$ global symmetry. 
The coordinate $z_1$ of the threebrane on the \cp\ corresponds to the gauge
invariant field $u=\tr(X^2)$ (where $X$ is the adjoint scalar in the
$SU(2)$ gauge multiplet). The $D_N$ singularity may be interpreted as
a ${\bf Z}_2$ orientifold point with $N$ mutually local sevenbranes, and the
$N$ quarks then arise from strings between the threebrane and these
sevenbranes \refs{\senetc,\brprobe}. In eight 
dimensions there is (in spacetime) an $SO(2N)$ adjoint
scalar which is part of the vector multiplet (corresponding
to the location of the sevenbranes on the $\bf{P^1}$), and on the brane probe
this corresponds to a mass matrix for the quarks. 

When we fiber this singularity over an additional \cp, the $SO(2N)$
adjoint scalars are lifted, though we expect them to still be visible
to a threebrane probe when it is far from any singular fibers, since in
the IR the probe will just see the eight dimensional theory.  At special
points on the bottom \cp\ the singularity becomes worse.   
In the F-theory description, these points correspond to
additional sevenbranes intersecting the original sevenbranes we had
(corresponding to the $D_N$ singularity). When the intersecting
sevenbranes are mutually local to the $N$ sevenbranes discussed above, we get
$SO(2N)$ charged matter in the $\bf 2N$ 
representation.\foot{When monodromies on the bottom \cp\ break
$SO(2N)$ to a smaller
group, this localization of matter does not occur.
Such cases will be discussed in the following subsection.}
In other cases more interesting representations may
appear.  
We will discuss these in the next section.

Particles which arise from strings between the threebrane and a
$(p,q)$ sevenbrane have electric charge $p$ and magnetic charge $q$ in the
threebrane field theory, so we can expect to have a local Lagrangian
description (which is different from the eight dimensional description)
only at points where we have massless $\bf{2N}$ fields localized in
spacetime. At such a
point (which we choose to be at $z_2=0$) the spacetime $\bf{2N}$s 
should correspond to parameters in the ${\bf 2N}$ representation of
$SO(2N)$ in the worldvolume
theory, and we expect to 
have additional massless fields coming from the
strings between the threebrane and the additional sevenbranes. 
Thus, the Lagrangian for a threebrane
near a massless vector point should be of the form
\eqn\vector{W = Q^a X Q^a + z_2 q^1 q^2 + \lambda_{ij} q^i X q^j 
+ m^a q^1 Q^a,}
where $Q^a$ ($a=1,\cdots,2N$), $q^i$ ($i=1,2$) are $SU(2)$ doublets,
$z_2$ is a worldvolume singlet field (corresponding to the location of
the brane in the ``bottom'' \cp), and $m^a$ is the spacetime 
$\bf{2N}$. The $ \lambda q X q$ term can be understood as follows.  The
discriminant of the elliptic fiber near the singularity looks
like
\eqn\discso{\Delta=z_1^{N+2}(\alpha z_2^2+\beta z_1+o(z_1^2))}
where $\alpha$ and $\beta$ are functions of $z_2$ which
go to constants as 
$z_2 \to 0$.\foot{Note that the formula
given for $\Delta$ on the $SO(8)$ locus
in \S4.7\ of \bikmsv\ is incorrect.  The correct
formula
is $\Delta \sim z_{1}^{6}(f_{2n+8}^{2} q_{n+4}^{2} + o(z_1))$
which agrees with \discso.}
Zeroes of the discriminant
correspond to sevenbranes which generically give rise to extra massless
fields on the threebrane; from \discso\ we see that this occurs
when $\alpha z_2^2+\beta z_1=0$.  This is reproduced by
our Lagrangian if we identify $\det(\lambda)$ with $\beta/\alpha$.
The last term in \vector\ reflects the Higgs mechanism in
spacetime, which corresponds to global symmetry breaking
on the brane.  We have chosen $q^1$ to be the field that 
couples to $m^a$ by an $SO(2)$ rotation of the $q$ fields.
When $m^{2N}\ne 0$, we see that the global
$SO(2N)$ symmetry is explicitly broken to $SO(2N-1)$ as required.

The Lagrangian written above (and any others appearing in this paper)
should be understood as giving a ``phenomenological'' description of
the low energy physics on the brane probe when it is near the
singularity. We expect the sort of terms we write, as well as terms
involving higher orders in $z_1$ and $z_2$ which we suppress, to arise
from interactions between the (fundamental) 
strings ending on the probe and on the
spacetime sevenbranes. At weak coupling it should be possible (for the
theories discussed in this section) to actually
compute these terms from string theory. 

 From the expression for the $j$ function \j, and from the known form
of the Seiberg-Witten curve, 
we see that 
for these theories $\tau\sim \log(z_1^{N-4}z_2^2)$, which
goes to $i\infty$ at the singularity ($z_1=z_2=m^a=0$).  This 
indicates that,
as expected, these theories are free in the infrared.
In particular, the theory \vector\ flows (for $N \ge 4$) to a free
${\cal N}=2$ theory with $SO(2N+2)$ global symmetry. 
This is in accord with the picture of Katz and Vafa \katz, which
describes the appearance of ${\bf{2N}}$ matter in the six dimensional
theory by starting with an $SO(2N+2)$ theory in eight dimensions and
turning on a $z_2$ dependent adjoint VEV 
which breaks $SO(2N+2)$ to $SO(2N)$.  
Both $z_2$ and
the mass parameters $m^a$ become part of the $SO(2N+2)$ adjoint
parameter of this infrared theory. Turning on either $z_2$ or $m$ is
equivalent to turning on the adjoint in the $SO(2N+2)$ theory, and
causes us to flow to the $SO(2N)$ ${\cal N}=2$ theory. 
When more than one ${\bf{2N}}$ field is present, for generic
values of $z_2$ the IR theory will have an $SO(2N)$ adjoint parameter
that includes a contribution of the form $h(z_2)m_i^{[a} m_j^{b]}$ where
$m_i$ and $m_j$ are any two spacetime ${\bf 2N}$ fields and
$h(z_2)$ is some function of $z_2$.\foot{The worldvolume Lagrangian description
of the theories with multiple ${\bf{2N}}$s in spacetime is provided below.} 
Turning on
two of these spacetime fields breaks the spacetime gauge group to 
$SO(2N-2)$, and this is realized by the adjoint breaking
in the worldvolume ${\cal N}=2$ superconformal theory.

There are special loci in the
moduli space of the compactification where $k$ spacetime
vector points come together.  In this case the discriminant goes
like 
\eqn\discvect{\Delta\sim z_1^{N+2}(\alpha z_2^{2k}+\beta z_1),}  
and we see that the point with extra massless matter
occurs at $\alpha z_2^{2k}+\beta z_1=0$.
The total number of spacetime
${\bf 2N}$ fields is $n+2N-4$, suggesting that
the Lagrangian \vector\ should be generalized to
\eqn\vectorII{W = Q^a X Q^a + P_{n+2N-4}(z_2) q^1 q^2 
+ \lambda_{ij} q^i X q^j 
+ \sum_{\alpha = 1}^{n+2N-4} (p_{\alpha}(z_2) m^a_\alpha q^1 + 
{\tilde p}_{\alpha}(z_2) {\tilde m}^a_{\alpha} q^2) Q^a} 
where $P_{n+2N-4}$ is a polynomial of
degree $n+2N-4$ in $z_2$ whose zeros correspond to the location of the
spacetime matter fields \bikmsv, $p_{\alpha}$ and ${\tilde
p}_{\alpha}$ are polynomials in $z_2$, and $m^a_{\alpha}$ and ${\tilde
m}^a_{\alpha}$ are the two complex scalars in the hypermultiplet
corresponding to the $\alpha$th massless ${\bf 2N}$ field.  
This has two effects.
First, when we bring $k$ zeroes of $P_{n+2N-4}$
together, we recover the behavior determined by the discriminant
\discvect.  Also, we have incorporated the fact that when
$k$ spacetime vector points converge, we should have $k$ spacetime
vectors which couple to naively distinct operators 
$z_2^i q^1 Q^a, z_2^i q^2 Q^a$ (for $i<k$ these are in the chiral ring).  
In this case the worldvolume
theories corresponding to different values of $k$ all
flow to the same (free) theory in the infrared, in which $z_2$ is a
free field.

\subsec{$SO(2N-1)$ Theories} 

The generic theory with a $D_N$ singularity at a point on the top
\cp\ actually yields $SO(2N-1)$ gauge symmetry in spacetime 
\refs{\aspg,\bikmsv}.\foot{Except in the case $N=4$ where $G_2$
is the generic symmetry and $SO(7)$ is a special case, or in cases
where the spacetime field content is not compatible with Higgsing to
$SO(2N-1)$.} 
At special points on the bottom \cp\ there is monodromy implementing
an outer automorphism on the $D_N$ Dynkin diagram, yielding a
$B_{N-1}$ diagram. 

In F-theory, the locus on which there is no such monodromy 
has discriminant
\eqn\DnDisc{\Delta \sim z_{1}^{N+2}(q_{n+2N-4}(z_2)^{2} + o(z_1)).} 
Breaking $SO(2N)$ to $SO(2N-1)$ involves generalizing $q_{n+2N-4}^{2}$ to a
generic polynomial $P$ of degree $2n+4N-8$ in $z_2$.  This means that,
for small values of $m^{a}$, the special 
$\bf{2N}$ points on the $z_2$ plane each
split into two points as one deforms from the $SO(2N)$ to the $SO(2N-1)$ 
theory.    

The monodromy is localized at the zeroes of $P$ on the $z_2$ plane. 
Therefore, at 
generic points, one expects to flow in the IR to a theory with $SO(2N)$
symmetry, while at the zeroes of $P$ sometimes the IR theory will 
exhibit only $SO(2N-1)$ symmetry.   
 From the Lagrangian \vectorII\ we see that if $m^a$ is nonzero, the
$SO(2N)$ symmetry is explicitly broken to $SO(2N-1)$.  However, there are
still $2N$ massless quarks in the theory, and the
theory will generically flow  
to the ${\cal N}=2$ supersymmetric theory with a full $SO(2N)$ global
symmetry, which is an IR-stable fixed line
\finite.  The coupling of the extra massless 
quark to $X$ generically flows to the ${\cal N}=2$ value in the infrared.

For $N > 4$ the theory is free in the IR and the meaning of
this statement is not clear.  For $N=4$, the theory is finite
at generic points, but still becomes free at the special points we are
discussing, as we will see below.  For $N=3$   
we find interesting behavior at the monodromy points, which we will
discuss 
in \S3.4.

Let us show how this behavior is reproduced by the Lagrangian
\vectorII.  Consider the case $N=4$ and $m\ne 0$.  Then, integrating
out the massive quarks, we obtain the effective superpotential
\eqn\vectorIII{W=\sum_{a=1}^7 Q^aXQ^a+
(z_2^2+\lambda_{12}^2z_1+\lambda_{22}m^2)qXq}
where $q$ is an appropriate combination of $q^2$ and $Q^8$
(i.e. the combination that remains massless).  Now, we
see that for nonzero $m$, there are two different types of
behavior, depending on the value of $z_2$.  For generic
$z_2$, the $qXq$ term is nonzero, and the theory will flow to the 
${\cal N}=2$ $SO(8)$ theory in the infrared.  However, for
$z_2$ a solution of $z_2^2+\lambda_{22}m^2=0$,
the bare $qXq$ coupling vanishes and it cannot be perturbatively
generated.  Thus, for these two special points on the
$z_2$ plane, one might expect to flow to a fixed point with
$SO(7)$ global symmetry and ${\cal N}=1$ supersymmetry. 
In fact, the theory flows to a free theory in the 
infrared, as can be seen by examining the behavior of 
the $j$ function \j\
at the special points.  This can also be seen
by looking at the beta functions
for the couplings in \vectorIII\ at the special points. 
For non-zero $z_1 \sim \tr(X^2)$, the $Q$ quarks are all
massive (as in the SW $N_f=4$ theory), but for $z_2$ such that the
coefficient of the $q X q$ term vanishes, $q$ will remain massless.
In this case
we will flow to an ${\cal N}=2$ $U(1)$ gauge theory with one 
massless electron.  
This is consistent with the form of the discriminant of the SW curve
around the $SO(7)$ monodromy points.
Note that the original
singularity at the $\bf 2N$ point has split as we expected, and that our
superpotential explicitly shows that the matter on the threebrane is no
longer localized at a specific value of $z_2$, just like the spacetime
matter in this case.

The local behavior of the discriminant at points at which a monodromy
breaks $SO(8)$ to $G_2$ is exactly the same as the behavior at $SO(7)$
monodromy points. Thus, even though we have no local Lagrangian
description of these theories, we expect that they will also flow to
free theories in the IR, though it is not clear which variables
(``electric'' or ``magnetic'') should become free in this case.

\subsec{$SU(4)\simeq SO(6)$}

So far we have only used the Lagrangian formulation
for cases where there were no quantum corrections to the classical
moduli space.  But, as discussed by Sen \senetc, 
F-theory
also correctly describes the quantum corrections to the
moduli space when quark masses are added to the $SO(8)$ theory.
In particular, giving a mass to one quark leads to
a theory with $SO(6)$ global symmetry.  This is the
$N_f=3$ case studied by Seiberg and Witten \swtwo.
The quantum moduli space of this theory has two
singularities, one of which corresponds to
a massless ${\bf 4}$ and the other to a massless singlet.
In F-theory we find an $A_3$ singularity at $z_1=0$, and
we identify this with the Seiberg-Witten ${\bf 4}$ point.  
The field theory of the probe
near such a singularity is an ${\cal N}=2$ $U(1)$ gauge theory
with four massless electrons.  

Special points on the $z_2$ plane are associated with
spacetime fields in the ${\bf 4}$ or ${\bf 6}$ representation
of $SO(6)$.  Consider the theory near a ${\bf 6}$ point:  we
expect the classical superpotential to be given by \vector.
When $m=z_2=0$, there are 4 massless quark hypermultiplets,
and the theory will flow in the IR to the $SO(8)$ ${\cal N}=2$ theory (with a
finite value of the gauge coupling).
At this fixed point, $z_2$ is the mass of one flavor.  The
curve around this point is exactly the same as the 
Seiberg-Witten curve for $N_f=4$ with one small mass turned
on \swtwo. In both cases the order $6$ singularity at the origin of moduli
space splits into one singularity of order $4$ (which we can keep at
$u\equiv\ z_1=0$) and two singularities of order $1$ which occur at $u
\sim z_2^2$.

\subsec{$SO(5) \simeq Sp(2)$ Fixed Lines}

As in the discussion of \S3.2, we can turn on a VEV for
the spacetime $\bf 6$ field. Again, the F-theory picture suggests the
existence of two special
values of $z_2$ (for $z_1=0$) for which the theory would flow to a
theory with $SO(5)$ global symmetry, and this is supported by a
classical analysis of our superpotential \vector. After integrating out
the massive fields we find
that near such a special value of $z_2$
the $SO(5)$ theory is described by the superpotential
\eqn\sofive{W \sim \sum_{a=1}^5 Q^a X Q^a + (z_2 + h z_1) q X q.}   
In this case our theory
is an asymptotically free theory (with one adjoint chiral multiplet
and $6$ doublets), so quantum corrections are expected to be important
(as they were in the $SO(6)$ theory). 

The form of the $j$ function \j\ as one approaches a monodromy 
point at $z_2=0$ is as follows: 
\eqn\jsofive{j(\tau)\sim {(b_1z_1^2+b_2z_1z_2+b_3z_2^2)^3
\over{z_1^4(c_1z_1^2+c_2z_1z_2+c_3z_2^2)}}.}
Approaching the fixed point with a given ratio
of $z_1$ to $z_2$, we see that the limiting value of
$\tau$ varies as a function of the moduli $b_i,c_i$ (which are related
to the couplings in the superpotential \sofive).
This suggests the existence of a fixed line
with every possible value of $j(\tau)$.  Indeed,
this behavior of the $j$ function near the singularity
is similar to that of the $j$ function near the singularity
corresponding to the ${\cal N}=2$ $SO(8)$ theory with
a small mass for one quark, which was discussed in \S3.3.  
In that case we know there is a fixed line.
From \jsofive\ we see that there is an ambiguity in the limiting
value of $\tau$ coming from the choice of 
$z_1/z_2$ as we approach the singularity.  This
reflects the fact that $z_1$ and $z_2$ control the
masses of mutually nonlocal states, and it
prevents us from reading off the value of $j(\tau)$
at the singularity.  
  
The superpotential for these theories at the monodromy points
looks like
\eqn\sofiveII{W = \lambda \sum_{a=1}^5 Q^a X Q^a + h ~\tr(X^2) q X q.}   
We can check for the plausibility of the existence of a fixed line
along the lines of \finite.  Computing the beta functions 
for the gauge coupling
$g$, and for the couplings $h$ and $\lambda$, 
one finds that they are proportional to
the scaling coefficients 
\eqn\bfunc{A_g= 2 + 5 \gamma_Q + \gamma_q + 4 \gamma_X}
\eqn\btwo{A_{\lambda} = {1\over 2} \gamma_X + \gamma_Q}
\eqn\bthree{A_{h} = 2 + {3\over 2} \gamma_X + \gamma_q.} 
In particular, we find that
\eqn\lindep{A_g = A_{h} + 5 A_{\lambda}.}
 From \lindep\ we see that there are only two independent scaling
coefficients for three variables, so generically we expect that 
a fixed line should exist in $(g,h,\lambda)$ space.  This is 
consistent with the computation of $j(\tau)$ from F-theory, and
lends credence to our use of the superpotential \sofiveII\
and in particular to our assumption that the $ h \tr(X^{2}) qXq$ term 
plays an important role in the infrared.

Note that the \sofiveII\ fixed line does not pass through 
any weak coupling regime where a perturbative analysis would be
reliable. The behavior of the $j$ function suggests that the fixed
line should exist for any value of $j(\tau)$, but this does not
necessarily mean that it passes through a region of weak ``electric''
coupling -- perhaps when $j(\tau)$ is large it is actually the
``magnetic'' gauge coupling that is weak. In any case,
for $\lambda, h$ small the theory certainly flows 
to strong gauge coupling. 

We have found that at $z_1=z_2=0$ the theory flows to a superconformal
fixed point. Next, we should try to compute the dimensions of the
operators at this point. As in \refs{\ntwo,\generaliz}, 
we can compute ratios of
dimensions of operators directly from the elliptic curve describing
the behavior of the gauge coupling near the fixed point. In the
$SO(5)$ case, this curve looks like
\eqn\sofivecurve{y^2 = x^3 + x(z_2^2 + z_2 z_1 + z_1^2 + \cdots) +
(z_2^3 + z_2^2 z_1 + z_2 z_1^2 + z_1^3 + \cdots)}
where $\cdots$ denote terms of higher order in $z_1$ and/or $z_2$, and
constant coefficients for all terms are suppressed.
Taking the leading terms in the curve to have equal dimension, we find
that $[z_1]=[z_2]$.  In the next paragraph, we will normalize the 
dimensions of the fields by using information present in the string
theory.  Even before doing so, we can check whether the fixed point
is ${\cal N}=2$ or ${\cal N}=1$ superconformal. 
If the theory $\it does$ have ${\cal N}=2$
supersymmetry, we can determine the dimensions 
(as in \refs{\ntwo,\generaliz}) by using
the fact that $\int u {dx \over y}$ should have dimension one (since
it gives the mass of BPS states). In our case, we identify $z_1$ with
$u$, and find that this assumption leads to $[z_1]=[z_2]=2$. While
this is the natural dimension for $z_1$ in an ${\cal N}=2$ theory (since
$z_1 \sim \tr(X^2)$ and $X$ is in a vector multiplet), it seems
inconsistent with ${\cal N}=2$ supersymmetry to have
$[z_2]=2$.  This is because $z_2$ has no ${\cal N}=2$ superpartner with
the same dimension, and hence would have to behave as a decoupled 
free field
in the infrared (i.e., have $[z_2]=1$) for 
the fixed point to have ${\cal N}=2$ supersymmetry. 
Thus, the ${\cal N}=2$ assumption seems to lead to a
contradiction, and we are led to believe that the new superconformal
theory we have found actually has only ${\cal N}=1$ supersymmetry. It is
clear that the global symmetry of this superconformal theory is at
least $SO(5)$. In principle, the global symmetry might be enhanced in
the infrared, but we see no reason for this to happen in this case.

In fact, using the information present in F-theory we can even 
precisely normalize
the dimensions $[z_1], [z_2]$ at our ${\cal N}=1$ fixed point.\foot{ 
We thank C. Vafa for suggesting this to us.}
The Calabi-Yau $M$, which we take to be elliptically fibered with
base $F_n$, has a holomorphic three-form $\Omega$.  It follows from
special geometry that, if the elliptic fiber is taken to have
constant volume,  
\eqn\vol{\int_M \Omega \wedge {\overline{\Omega}} \sim Vol(M) \sim
Vol(T^2) \times Vol(F_n)}
and that the volume form on the Coulomb branch of the field theory
(which we identify with the base $F_n$)
is given by
\eqn\ftmod{\int_{T^2} \Omega \wedge {\overline{\Omega}}~.} 
So, holding the volume of the $T^2$ fixed (as F-theory instructs
anyway), and performing a scale transformation in the field theory,
we expect $\Omega$ to have $[\Omega] = 2$.
A general formula for $\Omega$ given a threefold defined by an equation
$W(x,y,z_1,z_2)=0$ is
\eqn\omegform{\Omega ~\sim~ {{dx \wedge dz_1 \wedge dz_2}\over {{\partial W}
\over {\partial y}}}~.} 
 From \sofivecurve, we see that $\Omega$ is given in a neighborhood
of the singularity as
\eqn\omega{\Omega ~\sim~{{dx\wedge dz_1 \wedge dz_2}\over y}~.} 
Hence,
\eqn\dimconst{[z_1] + [z_2] + [x] - [y] = 2,}
and using the known ratios between the dimensions we find that
\eqn\normdim{[z_1] = [z_2] = {4\over 3}~.}
Note that if there is a fixed line, the dimensions do not
vary along it.  This might indicate that the photon, whose
gauge coupling changes along the fixed line, decouples
from the nontrivial conformal theory.  

The realization of our quantum field theory on
a threebrane in F-theory enables us to compute the
K\"ahler potential even though the theory only has ${\cal N}=1$
supersymmetry.  The metric on the Coulomb branch of our
field theory is just the restriction of the elliptic Calabi-Yau
metric to the $F_n$ base.  Although the Calabi-Yau metric
is not known in general, it can be computed in the vicinity
of singularities (for example near conifold singularities
the metric is given in \conmet).  From the conformal field
theory point of view it is only this information that is
relevant in any case.  The global form of the metric depends
on massive degrees of freedom that decouple from the
fixed point theory.  It would be interesting to try to
understand the appearance of \omega\ directly in the field theory.

As described above, we can flow to the new $SO(5)$ superconformal
theory by starting with the $SO(6)$ theory at a massless ${\bf 6}$
point, and turning on $z_2$ and $m$ with an appropriate ratio between
them. At $z_2=m=0$, that theory flows in the infrared to the ${\cal
N}=2$ $SU(2)$ gauge theory with $N_f=4$, which has an $SO(8)$ global
symmetry and a parameter in the adjoint of $SO(8)$ (corresponding to a
mass matrix for the quarks). The $N_f=4$ theory also has (in general)
additional
parameters, such as the Yukawa coupling appearing in front of a term
$Q^a X Q^b$ in the superpotential, which is in the symmetric tensor
representation of $SO(8)$. In terms of the infrared conformal theory,
turning on $z_2$ and/or $m$ corresponds to turning on both adjoint
perturbations (which preserve ${\cal N}=2$) and symmetric tensor
perturbations (which break ${\cal N}=2$). For generic values of the
parameters the theory will flow back to an ${\cal
N}=2$ theory in the infrared, despite the explicit breaking we turned on. 
However, for
special values of $z_2$ and $m$ (as described above), 
we will actually flow to the ${\cal N}=1$ $SO(5)$ theory.

We can also discuss what happens when we bring $k$ monodromy points
together. For even $k$, we expect at least the $SO(6)$ symmetry to be
restored in this case (since each point corresponds to a ${\bf Z}_2$
monodromy), and presumably the behavior is similar to the theories
discussed in the previous subsection (i.e. the theory flows to the
${\cal N}=2$ theory with $SO(8)$ global symmetry). For odd $k$ the behavior
of the theory is not a priori clear, and we have not been able to uniquely
identify a Lagrangian that will flow to these points. 
Performing the calculation \dimconst\ for this case yields
$[z_1]={4k\over{k+2}}$ and $[z_2]={4\over{k+2}}$.  
For $k=2$ this is consistent with the expectation that
the theory flows to the ${\cal  N}=2$ $SO(8)$ theory.
For higher $k$, the naive dimension for $z_2$ is less than
one.  This would seem to indicate that the theory
is nonunitary, since the superconformal algebra requires
scalar fields in a unitary theory to have dimension $\ge 1$. 

However, as discussed in \refs{\ksw,\ntwo}, the proper interpretation
of this is that $z_2$ is actually a free field which
enters the Lagrangian as part of a ``dangerously irrelevant''
operator.  Specifically, the Lagrangian includes terms of the form
$\mu^{-\delta} \int d^{2}\theta z_2 {\cal O}$, where
the dimension of the operator ${\cal O}$ satisfies $3>[{\cal O}]=2+\delta>2$
and $z_2$ is a free field.
At the fixed point, this operator is irrelevant and
flows away in the infrared.  However, upon giving a vacuum
expectation value to $z_2$, the resulting deformation
$\mu^{-\delta} \int d^{2}\theta {\langle z_2 \rangle}{\cal O}$ becomes
relevant.  Now we can identify $[z_2]$ computed from
the curve with the dimension of $\mu^{-\delta} {\langle z_2\rangle}$.

Going back to our example, we learn that for $k>2$ the
physical dimension of $z_2$ is one, and it decouples
from the theory.  Plugging this back into \dimconst, we
recover the formula $[z_1]+[x]-[y]=1$, which is
just the ${\cal N}=2$ formula.  This is not surprising
since when $z_2$ decouples we expect the theory to
reduce to an appropriate ${\cal N}=2$ brane theory
in eight dimensions. Calculating the dimension of $z_1$, we find that
for all $k > 1$, $[z_1] = 2$.
For even $k$, this is consistent
with our expectation that these theories flow to the
${\cal N}=2$ $SO(8)$ theory.  For odd $k$ the situation is
less clear:  the dimensions are 
those of the ${\cal N}=2$ $SO(8)$ theory, but on the other
hand we would expect the monodromy breaking 
$SO(6)$ to $SO(5)$ to reduce the global symmetry
here just as in the $k=1$ case \sofive.

\newsec{Exceptional Groups and Other Nontrivial Fixed Points} 

In this section we will discuss fixed points for which we have no
local Lagrangian description. For threebrane probes at $E_n$
singularities no Lagrangian description is known already in the eight 
dimensional case, and we will discuss the ${\cal N}=1$ superconformal
theories related to these singularities in \S4.1 and \S4.2. 
In \S4.3  we discuss additional
examples of such theories, which are
obtained by probing points at which spacetime fields in
the spinor of $SO(2N)$ are localized.  In analogy to the way ${\bf
2N}$s in spacetime coupled to operators in the ${\bf 2N}$
representation in the worldvolume, we expect couplings of spacetime
fields in the spinor of $SO(2N)$ to worldvolume operators in the
spinor of $SO(2N)$.

In the $SO(8)$ case there is a triality symmetry between 
vectors and the two types of spinor representations. 
Thus, in this case, when we approach a spinor point an additional
``magnetic'' quark becomes massless and the superpotential 
in terms of magnetic variables is the same as above \vectorII.
In this case of course these points will just be
weakly coupled in terms of the ``magnetic'' variables.
In the more general situation, for example $SO(10)$, we
still expect to find additional massless magnetically charged
particles at these points.
In this case the $Q^a$, which tranform in the ${\bf 10}$
representation, are necessarily ``electric'' variables, and
no local Lagrangian description including both ``electrically'' and
``magnetically'' charged particles is known.  
This is similar to the fixed points discussed in \refs{\argdoug,\ntwo}.
In general, when the threebrane approaches a point
where a spacetime field in the representation 
${\bf R}$ is localized, additional degrees of freedom
will come down which are parts of composite operators
on the worldvolume which are in the ${\bf \bar R}$ representation.  
We expect the spacetime fields, which are parameters
on the brane, to couple to the worldvolume ${\bf \bar R}$s. Then,
when the spacetime fields obtain VEVs,
the global symmetry breaks appropriately as discussed
above.  

\subsec{Evidence for $F_4$ Theory}

Let us begin by considering what happens when the upper 
\cp\ develops an $E_6$ singularity.  The curve
in this case is of the form
\eqn\Esixcurve{y^2=x^3+(z_1^3f_{n+8}(z_2)+\cdots)x+z_1^4g_{2n+12}(z_2)
+\cdots}
and the discriminant looks like
\eqn\Esixdisc{\Delta\sim z_1^8g_{2n+12}^2+\cdots.}
In \Esixcurve\ and \Esixdisc, $\cdots$ represents terms
of higher order in $z_1$.  For generic
$g_{2n+12}$, the spacetime gauge symmetry is actually
$F_4$.  This occurs, as explained in \aspg\ and \bikmsv,
because of a ${\bf Z}_2$ monodromy about zeroes of $g_{2n+12}$ on
the lower \cp.  When $g_{2n+12}\sim q_{n+6}^2$,
the monodromies vanish and the spacetime gauge
symmetry is the full $E_6$.  In both cases, at generic values of $z_2$, the
threebrane theory flows to
the ${\cal N}=2$ $E_6$ theory of \refs{\seiberg,\Esix} 
by the adiabatic argument.  

Let us look at what happens when the probe approaches
a zero of $g_{2n+12}$ of multiplicity $k$.  Consider
first the cases with $k$ odd.  
Going around such points induces a monodromy on the $E_6$
Dynkin diagram which breaks $E_6 \rightarrow F_4$. 
Therefore we expect
that $F_4$ is the global symmetry in the infrared.
This is analogous to the $SO(5)$ global symmetry
we found in \S3.4, when we had a Lagrangian description.

Given the curve \Esixcurve, we can compute 
the dimensions of $z_1$ and $z_2$ at the fixed point
as a function of $k$, as in the previous section. 
To do this, let us take
all leading terms in the curve to have the same dimension.
This leads to the result
\eqn\Esixrat{[z_1]={8k\over{k+2}},~[z_2]={4\over{k+2}}.}

For $k=1$, we get a nontrivial interacting fixed
point which we expect to have $F_4$ global symmetry.
This theory cannot have extended supersymmetry since
$[z_2]=4/3$.  
Computing the value of the $U(1)$ gauge coupling constant on
the Coulomb branch near our singularity using the $j$-function
\j, we find $\tau = i$.

For even $k$, we can describe this singularity by starting with an
$E_7$ singularity in the upper \cp\ and giving the adjoint a
$z_2$-dependent VEV \katz,
suggesting that 
the theory flows to the $E_7$ ${\cal N}=2$ theory in the infrared. 
The dimensions we find are consistent with this.
 From this theory we can flow to the $k=1$ theory
by turning on a ${\bf 27}$ field in spacetime, which corresponds to
splitting the two zeros of $g_{2n+12}$. In the ${\cal N}=2$
superconformal theory this would correspond to turning on
perturbations which break ${\cal N}=2$ (these exist in addition to the
adjoint perturbations that preserve ${\cal N}=2$) and which break
$E_7$ to $F_4$. This is analogous to the similar situation discussed
in \S3.4.
For odd $k>1$, the dimensions are the same as for even $k$, but, 
as in the $SO(5)$ case, the
global symmetry (and the number of supersymmetries) is not clear.

\subsec{New $E_7$ Fixed Point}

Similarly, we can go to the $E_7$ locus on which the curve looks like
\eqn\esevc{y^2 = x^3 + (z_1^3 f_{n+8}(z_2)+ \cdots) x + z_{1}^5 g_{n+12}(z_2)
+ \cdots,}
and the discriminant is 
\eqn\esevdisc{\Delta \sim z_1^{9} (f^{3}_{n+8}(z_2) + o(z_1)).} 
At generic points in the $z_2$ plane, the theory flows (by the adiabatic
argument) to the previously studied ${\cal N}=2$ $E_7$ superconformal theory.
One can also examine the theory in the neighborhood of a zero of
$f_{n+8}(z_2)$ of order $k$.  Unlike the previous cases, here at
a zero of order $k$ one obtains $k$ $\it half$-hypermultiplets
(in the $\bf 56$ representation).  For $k=1$, and in general when one
half-hypermultiplet is associated with a point in other cases,
we will argue that
the theory in the infrared is a new ${\cal N}=1$ fixed point. 

At such a point, the curve \esevc\ locally reduces to
\eqn\specpt{y^2 = x^3 + z_1^{3} z_{2}^k x + z_{1}^5}
We assume that all terms in the curve \specpt\ are equally
relevant. Then, the by now familiar computation of dimensions
gives $[z_1]={12k\over{k+2}}$ and $[z_2]={4\over{k+2}}$.  
For $k=1$, we again obtain a new ${\cal N}=1$ superconformal
fixed point.  The global symmetry group contains $E_7$, and
the gauge coupling on the Coulomb branch as one approaches the
singularity is $\tau = e^{{2\pi i}\over 3}$. 

For $k>1$ and odd, the issues are much the same as the odd $k$, $k>1$ case
of \S4.1.
For $k$ even it seems reasonable, from the description of Katz and Vafa
\katz, to conjecture that these theories flow to an 
${\cal N}=2$ $E_8$ fixed point. Note that no similar picture exists to
describe matter in half-hypermultiplets, so it is reasonable to assume
that the behavior of the brane probe would be different in these
cases.

\subsec{SO(K) at Spinor Points}  

One can analyze $SO(K)$ theories at various special points 
by the same methods we have been using in previous sections.
One sees from \bikmsv\ that half spinor points occur on the $SO(11)$
and $SO(12)$ loci.  By approaching such points, we expect to obtain
new ${\cal N}=1$ superconformal theories whose global symmetry is at least
$SO(12)$ (and may be enhanced to $E_7$).  For the
$SO(12)$ theory at a ${1\over 2}{\bf 32}$ point, we find
dimensions $[z_1]=8/3$ and $[z_2]=4/3$.
The coupling $\tau$ approaches $\tau = i$ near the singularity.

For the other $SO(K)$ theories (or for the $SO(11)$ and $SO(12)$ cases
when we merge an even number of half spinor points), we expect to get
${\cal N}=2$ theories as we approach points where spinors are
localized. The computation of
scaling dimensions gives results consistent with this expectation.
For example, on the $SO(10)$ locus by approaching
$\bf {16}$ points one probably
flows to the $E_6$ ${\cal N}=2$ superconformal theory.
On the $SO(8)$ locus at spinor points one flows to a (free) theory with
$SO(10)$ global symmetry (the $SU(2)$ gauge theory with $N_f=5$).

We can also consider what happens when an $SO(K)$ spinor point
approaches a vector point.  Let us do the analysis for $SO(10)$ here.
The curve at a vector+spinor point is
\eqn\vectspin{y^2=x^3+x(z_1^2z_2^2+z_1^3+\cdots)+
(z_1^3z_2^3+z_1^4z_2+z_1^5+\cdots)}
where here $\cdots$ indicates terms of higher order in
$z_1,z_2$.  Computing the dimensions,
we find $[z_1]=8/3$ and $[z_2]=4/3$.  
Thus, this must also be a new nontrivial ${\cal N}=1$ fixed point.
Unlike the case with a single localized vector multiplet,
there is no direct D-brane interpretation of the matter
in this case along the lines of \katz.
Other similar interesting possibilities exist but we will
not discuss them here.

\newsec{Discussion}

By examining the behavior of threebrane probes in F-theory on elliptic 
Calabi-Yau threefolds,
we have found evidence for the existence of many new nontrivial ${\cal N}=1$
supersymmetric renormalization group fixed points.  In five cases we can
actually prove, using the fact that $z_2$ is not a free
field, that the theory in the infrared has only ${\cal N}=1$ supersymmetry.
It is natural to think that in other cases (obtained by colliding
various singularities) there are also new ${\cal N}=1$ fixed points.
The five new fixed points we found which are $\it definitely$ only ${\cal N}=1$
superconformal are:
\medskip
\noindent
1) The theory with $SO(5)$ global symmetry group, obtained by approaching
a monodromy point (on the $z_2$ plane) on the $SO(5)$ locus in F-theory.

\noindent
2) The theory with $F_4$ global symmetry, obtained by approaching a monodromy
point on the $F_4$ locus in F-theory.

\noindent
3) The theory with $E_7$ global symmetry which arises near a half $\bf{56}$
point on the $E_7$ locus in F-theory.

\noindent
4) The theory obtained by approaching a half spinor point on the $SO(12)$
locus.

\noindent
5) The theory obtained by having a brane probe approach 
a vector + spinor point on the $SO(10)$ locus. 
\medskip
\noindent
It is possible 
that in many cases we discussed here
where the dimensions are consistent with ${\cal N}=2$ supersymmetry, 
the theory nevertheless has only ${\cal N}=1$ supersymmetry 
in the IR (or flows to a new
${\cal N}=2$ fixed point). It is natural to conjecture that a systematic
classification of all ${\cal N}=1$ superconformal theories with
Coulomb phases by algebraic methods (such as the ones we used) may be
possible.

It is interesting to note that the cases for which we have been able
to rule out the possibility of ${\cal N}=2$ supersymmetry are exactly
the cases for which we cannot interpret the worsening of the
singularity as arising from a $z_2$-dependent VEV for a spacetime
field, as described in \katz. Thus, it is natural to speculate that
whenever such a description does exist, the theory on the probe will
actually flow to an ${\cal N}=2$ superconformal theory (with the
appropriate enhanced global symmetry). The dimension of $z_2$ when we
have such a description always comes out to equal one, so it can
naturally become part of an adjoint ``mass'' parameter of the theory
with the larger global symmetry.

It would be interesting to understand the Higgs branches of the
theories we have found.  We expect the Higgs branch of the theories
with global symmetry $G$ to correspond to (or at least to include)
the moduli space of $G$-instantons 
on the seven branes.  In the cases with Lagrangian
descriptions, the Higgs phase corresponds to the space of
$Q$ and $q$ VEVs modulo the D-term and superpotential 
constraints.  Note that the superpotential may contain terms
which lift part or all of the Higgs branch, without effecting
our Coulomb branch analysis. 
Higher dimensional operators which have this 
effect may even be present at tree level in the field theory, and it
is difficult to determine their exact form since we only have ${\cal
N} = 1$ supersymmetry.

These or similar ${\cal N}=1$ superconformal theories may be relevant
for understanding what happens at singularities of 4d ${\cal N}=1$
supersymmetric string or F-theory compactifications.  For example,
$SO(32)$ strings on Calabi-Yau threefolds manifest a nonperturbative
$SU(2)$ gauge group at certain singularities in the moduli space;
nonperturbative dynamics in the $SU(2)$ explains the physics of the
singularities \kss.  It is natural to speculate that some classes of
singularities of $E_8 \times E_8$ strings on Calabi-Yau threefolds
will be explained by the appearance of a nontrivial fixed point
theory.\foot{We thank T. Banks for suggesting this possibility to us.}

It would also be interesting to study brane probes obtained by
wrapping a fivebrane on a higher genus curve in $K3$ compactifications
of the heterotic string.  At least for genus 2 the worldbrane quantum
field theory 
should have interesting dynamics.
Similarly, one could examine the behavior of $N$ coincident threebrane
probes \asty, 
which should manifest an $Sp(N)$ gauge theory on the worldvolume.
At special points in the Coulomb branch of this theory, there may be
new superconformal fixed points. More complicated singularities can
also occur in F-theory compactifications \berjoh. The study of
threebrane probes at these singularities should be interesting in its
own right \asty, and may also improve our understanding of the behavior of
the spacetime theory in these cases.

\medskip
\centerline{\bf{Acknowledgements}}

This work was supported in part by DOE grant DE-FG02-96ER40559.  
We are very grateful to Nati Seiberg for many valuable discussions, 
and to Cumrun Vafa for suggesting a method of normalizing dimensions.
We would also like to thank 
Paul Aspinwall, 
Tom Banks, Michael Douglas, and Steve Shenker for helpful comments.

\listrefs
\end